\documentclass[12pt]{article}
\usepackage{amsmath}
\usepackage{amsfonts}
\usepackage[english]{babel}
\usepackage{amssymb}
\usepackage{amsthm}
\usepackage{graphicx}
\usepackage[a4paper]{geometry}
\usepackage{hyperref}
\usepackage{amstext}
\usepackage{caption}
\usepackage{etoolbox}
\usepackage{makeidx}
\usepackage{amsfonts}
\usepackage{authblk} 
\usepackage{sectsty}
\usepackage{amscd}
\usepackage{mathrsfs}
\usepackage{dsfont}
\usepackage{enumitem} 
\usepackage[]{latexsym}
\usepackage{braket}
\usepackage{caption}

\usepackage{xcolor}

\newcommand{\be}{\begin{equation}}
\newcommand{\ee}{\end{equation}}
\newcommand{\bea}{\begin{eqnarray}}
\newcommand{\eea}{\end{eqnarray}}
\newcommand{\vsp}{\vspace{0.4cm}}

\newcommand{\appa}{\mathcal{A}}

\newcommand{\Tr}{\textit{Tr}}
\newcommand{\stsp}{\mathcal{S}}

\newcommand{\obsp}{\mathfrak{O}}

\newcommand{\gr}{\mathrm{g}}

\newtheorem{remark}{Remark}

\newtheorem*{proof*}{Proof}

\title{Geometrical Structures for Classical and Quantum Probability Spaces}
\author{Florio M. Ciaglia$^{1,2}$ and Alberto Ibort$^{3,4}$  and Giuseppe Marmo$^{1,2}$ \\
$^{1}$  Dipartimento di Fisica, Universit\`a di Napoli ``Federico II",\\ Via Cinthia Edificio 6, I-80126 Napoli, Italy  \\ $^{2}$ INFN-Sezione di Napoli, Via Cinthia Edificio 6, I-80126 Napoli, Italy \\ $^3$Departamento de Matem\'{a}ticas, Universidad Carlos III de Madrid\\
Avda. de la Universidad 30, 28911 Legan\'{e}s, Madrid, Spain.\\ $^4$ ICMAT, Instituto de Ciencias Matem\'{a}ticas (CSIC - UAM - UC3M - UCM) \\  
Nicol\'{a}s Cabrera,13–15, Campus de Cantoblanco, UAM, 28049, Madrid, Spain} 

\date{}

\begin{document}

\maketitle              

\begin{abstract}
On the affine space containing the space $\stsp$ of quantum states of finite-dimensional systems there are contravariant tensor fields by means of which it is possible to define Hamiltonian and gradient vector fields encoding relevant geometrical properties of $\stsp$.
Guided by Dirac's analogy principle, we will use them as inspiration to define  contravariant tensor fields, Hamiltonian and gradient vector fields on the affine space containing the space of fair probability distributions on a finite sample space and analyse their geometrical properties.
 Most of our considerations will be dealt with for the simple example of a three-level system. 
\end{abstract}

\section{Introduction}

The description of any physical system requires the identification of states, observables, a pairing between states and observables (for instance, expectation value functions), a dynamical law (evolution), and a composition rule for composite systems.
The main difference between classical and quantum descriptions is the use of a commutative/noncommutative  algebra containing the observables, and probability distributions/probability amplitudes, respectively (for classical and quantum descriptions).
A probabilistic interpretation of quantum mechanics was proposed by Born out of a detailed discussion of scattering problems.
In this interpretation, wave functions, defined on some ``configuration space'', may be interpreted as probability amplitudes because $\psi^{*}\,\psi$ gives a probability density, the amplitude point of view was required for a proper description of interference phenomena.
In abstract terms, out of an abstract Hilbert space $\mathcal{H}$, vectors are represented as ``square integrable functions'' on the joint spectrum of a maximally commuting set of operators.
When the joint spectrum turns out to be a differential manifold, it can be thought of as a ``sample space'' on which probability densities, built out of the wave functions, are defined.
From the point of view of the algebraic description of quantum mechanics, probability amplitudes are associated with the noncommutative structure of the algebra of observables while the ``sample space'', and probability distributions thereof, is associated with an Abelian algebra of observables.
For these reasons, often one speaks of quantum and classical probability theories, respectively.
As the quantum description contains more information than the classical one, it would be more appropriate to consider a ``dequantization'' procedure instead of a ``quantization'' one, i.e., a quantum-to-classical transition procedure.

Starting with the geometrical description of quantum mechanics, it is therefore interesting to consider the analogies and differences arising when dealing with amplitudes or probabilities in connection with the geometrical structures available directly on the space of quantum states with respect to those available on the classical side.
In general, thanks to the influence of Information Geometry, it is conventional to consider a Riemannian structure on the space of probability distributions along with dualistic connections.
Specifically, Cencov's theorem states that there is a  unique (up to a multiplicative factor) metric tensor on the space of probability distributions which is equivariant with respect to the category of Markov maps used in classical probability theory.\cite{cencov-statistical_decision_rules_and_optimal_inference}
This is known as the Fisher-Rao metric tensor.
Dual connections are then defined precisely with respect to the Fisher-Rao metric tensor.\cite{amari_nagaoka-methods_of_information_geometry}
The importance of the Fisher-Rao metric tensor in classical information geometry has motivated the search for a quantum analogue for such metric on the space of quantum states.
In Ref.\cite{petz-monotone_metrics_on_matrix_spaces} it is shown that in the quantum case there is an infinite number of metric tensors providing a meaningful generalization of the classical Fisher-Rao metric tensor.
    
In this paper we want to follow a sort of opposite path.
We would like to show that there are natural additional structures on the space of probablity distributions which allow to introduce a Poisson structure on a space of functionals which include physical observables.
This skew bracket, along with a possible symmetric bracket, would   parallel the structures available on the space of quantum states. 
Thus, in some sense, we merging the viewpoints expressed in our previous papers Refs.\cite{ciaglia_dicosmo_laudato_marmo-differential_calculus_on_manifolds_with_boundary.applications, ciaglia_dicosmo_ibort_laudato_marmo-dynamical_vector_fields_on_the_manifold_of_quantum_states, ciaglia_dicosmo_ibort__marmo-dynamical_aspects_in_the_quantizer-dequantizer_formalism}   in the probabilistic setting.
We shall briefly review first the quantum setting with their geometrical structures, then we consider the geometrical structures which may be defined on the space of fair probability distributions.
Most of our considerations will be dealt with for the simple example of a three-level system.

\section{Quantum states}\label{sec: quantum}

Here we will consider only a finite-level quantum system, that is, a quantum system whose Hilbert space $\mathcal{H}$ is such that $\mathrm{dim}(\mathcal{H})=n<\infty$.
The $C^{*}$-algebra of the quantum system is then the $C^{*}$-algebra $\appa\equiv\mathcal{B}(\mathcal{H})$ of all linear operators on $\mathcal{H}$.
The space $\obsp$ of observables is the real subspace of Hermitean elements in $\appa$.
The space of quantum states is the convex body $\stsp$ in the dual space $\obsp^{*}$ characterized by:

\be
\stsp:=\left\{\omega\in\obsp^{*}\colon \omega(\mathbf{A}\,\mathbf{A}^{\dagger})\geq0\;\;\forall \mathbf{A}\in\appa\,,\;\;\;\;\omega(\mathbb{I})=1\right\}\,.
\ee
For our purposes, we may exploit the fact that in finite-dimensions $\obsp$ and $\obsp^{*}$ are isomorphic, and thus we may indentify $\stsp$ with the set of density operators in $\obsp$, that is, all those positive semidefinite Hermitean operators $\rho\in\obsp$ such that $\Tr(\rho)=1$.

In  Refs.\cite{carinena_clemente-gallardo_jover-galtier_marmo-tensorial_dynamics_on_the_space_of_quantum_states, ciaglia_dicosmo_ibort__marmo-dynamical_aspects_in_the_quantizer-dequantizer_formalism, ciaglia_dicosmo_laudato_marmo-differential_calculus_on_manifolds_with_boundary.applications, ciaglia_dicosmo_ibort_laudato_marmo-dynamical_vector_fields_on_the_manifold_of_quantum_states} the Jordan-Lie algebra structure of $\obsp$ is exploited to introduce contravariant tensor fields on $\obsp^{*}$.
Specifically, if $f_{\mathbf{a}}$ denotes the linear function on $\obsp^{*}$ which is associated to $\mathbf{a}\in\obsp$ by means of $f_{\mathbf{a}}(\xi):=\xi(\mathbf{a})$, we have\footnote{Recall that, on every linear manifold $M$, the differentials of the linear functions  generate the cotangent space of $M$ at every point, therefore, in order for a tensor field to be uniquely defined, it suffices to evaluate  it on linear functions.}:

\be\label{eqn: tildelambda 1}
\Lambda(\mathrm{d}f_{\mathbf{a}}\,,\mathrm{d}f_{\mathbf{b}})=f_{\frac{\imath[\mathbf{a}\,,\mathbf{b}]}{2}}\,,
\ee
\be\label{eqn: G 1}
G(\mathrm{d}f_{\mathbf{a}}\,,\mathrm{d}f_{\mathbf{b}})=f_{\frac{\{\mathbf{a}\,,\mathbf{b}\}}{2}}\,,
\ee
where $[\,,]$ and $\{\,,\}$ denote, respectively, the matrix commutator and the matrix anticommutator.
Selecting an orthonormal basis $\{\mathbf{e}^{\mu}\}_{\mu=0,...n^{2}-1}$ in $\obsp$ such that $\mathbf{e}^{0}=\frac{1}{\sqrt{n}}\mathbb{I}$ with $\mathbb{I}$ the identity operator in $\mathcal{B}(\mathcal{H})$, we can introduce a Cartesian coordinate system $\{x^{\mu}\}_{\mu=0,...n^{2}-1}$ on $\obsp^{*}$ by setting:

\be
x^{\mu}(\xi):=\xi(\mathbf{e}^{\mu})\,.
\ee
In this coordinate system, we have:

\be\label{eqn: tildelambda 2}
\Lambda = c^{\mu\nu}_{\sigma}x^{\sigma}\,\frac{\partial}{\partial x^{\mu}}\,\wedge\,\frac{\partial}{\partial x^{\nu}}\,,
\ee
\be\label{eqn: G 2}
G= d^{\mu\nu}_{\sigma}x^{\sigma}\,\frac{\partial}{\partial x^{\mu}}\,\otimes\,\frac{\partial}{\partial x^{\nu}}\,.
\ee
The coefficients $c^{\mu\nu}_{\sigma}$ are the structure constants of the Lie product $\frac{\imath}{2}[\cdot\,,\cdot]$ in $\obsp$.
Note that, in the situation we are dealing with, $2c^{\mu\nu}_{\sigma}$ are the structure constants of the Lie algebra $\mathfrak{u}(\mathcal{H})$ of the unitary group $\mathcal{U}(\mathcal{H})$, and thus, they are  antisymmetric in all indices.
Analogously, the coefficients $d^{\mu\nu}_{\sigma}$ are   the structure constants of the Jordan product $\frac{1}{2}\{\cdot\,,\cdot\}$ in $\obsp$.
The structure constants $d^{\mu\nu}_{\sigma}$ are  symmetric in $\mu,\nu$, and we have $d^{0 0}_{j}=0$ and $d^{\mu\nu}_{0}=\frac{\delta^{\mu\nu}}{\sqrt{n}}$.
Furthermore,  the structure constants of both tensors are invariant with respect to unitary transformations, that is, the structure constants of the basis $\{\mathbf{e}^{\mu}\}$ equal those of the basis $\{\mathbf{e}'^{\mu}\}$, where $\mathbf{e}'^{\mu}=\mathbf{U}\,\mathbf{e}^{\mu}\,\mathbf{U}^{\dagger}$ with $\mathbf{U}\mathbf{U}^{\dagger}=\mathbb{I}$.

The Hamiltonian vector fields

\be\label{eqn: hamiltonian vector fields tildelambda}
\mathbb{X}_{\mathbf{a}}:=\Lambda(\mathrm{d}f_{\mathbf{a}}\,,\cdot)
\ee
associated with $\mathbf{a}\in\obsp$ by means of $\Lambda$, and the gradient-like vector fields:

\be\label{enq: gradient-like vector fields G}
Y_{\mathbf{b}}:=G(\mathrm{d}f_{\mathbf{b}}\,,\cdot)
\ee
associated with $\mathbf{b}\in\obsp$ by means of $G$, close on a realization of the Lie algebra $\mathfrak{gl}(\mathcal{H})$ of the Lie group $GL(\mathcal{H})$.
This realization integrates to a group action.
Specifically, if $\xi$ denotes the Hermitean operator uniquely associated with the linear functional $\omega\in\obsp^{*}$, the action of $GL(\mathcal{H})$ may be written as:

\be
\xi\mapsto \gr\,\xi\,\gr^{\dagger}= \mathrm{e}^{\imath\mathbf{a} + \mathbf{b}}\,\xi\, \mathrm{e}^{-\imath\mathbf{a} + \mathbf{b}}\,,
\ee
where $\mathrm{e}^{\imath\mathbf{a} + \mathbf{b}}=\gr\in GL(\mathcal{H})$.
It turns out that Hamiltonian vector fields are tangent to the affine subspace $\mathfrak{T}_{1}\subset\obsp^{*}$ of trace-one positive linear functionals of which $\stsp$ is a convex body, while  gradient-like vector fields are not.
Consequently, we decided to modify the symmetric tensor $G$ in such a way that the new gradient-like vector fields become tangent to $\mathfrak{T}_{1}$.
Specifically, the action of the new symmetric tensor on linear functions is:

\be
\mathcal{R}(\mathrm{d}f_{\mathbf{a}}\,,\mathrm{d}f_{\mathbf{b}})=f_{\frac{\{\mathbf{a}\,,\mathbf{b}\}}{2}} - f_{\mathbf{a}}\,f_{\mathbf{b}}\,,
\ee
In particular, when $\mathbf{a}=\mathbf{b}$, this expression gives the variance of $\mathbf{a}$.
The coordinate expression of $\mathcal{R}$ is:

\be\label{eqn: R}
\mathcal{R}= d^{\mu\nu}_{\sigma}x^{\sigma}\,\frac{\partial}{\partial x^{\mu}}\,\otimes\,\frac{\partial}{\partial x^{\nu}} - x^{\mu}\,x^{\nu}\,\frac{\partial}{\partial x^{\mu}}\,\otimes\,\frac{\partial}{\partial x^{\nu}}\,.
\ee
Gradient-like vector fields defined by means of $\mathcal{R}$ will be denoted as $\mathbb{Y}_{f}$.
Now, Hamiltonian vector fields $\mathbb{X}_{\mathbf{a}}$ and gradient-like vector fields $\mathbb{Y}_{\mathbf{b}}$ associated with traceless elements $\mathbf{a},\mathbf{b}\in\obsp$ close on a realization of the Lie algebra $\mathfrak{sl}(\mathcal{H})$ of the special linear group $SL(\mathcal{H})$.
Furthermore, they are all tangent to the affine hyperplane $\mathfrak{T}_{1}\subset\obsp^{*}$ of trace-one positive linear functionals\footnote{They are actually tangent to every affine hyperplane which is parallel to $\mathfrak{T}_{1}$.}.
However, contrary to the situation before, it turns out that the realization of $\mathfrak{sl}(\mathcal{H})$ given by Hamiltonian and gradient-like vector fields does not integrate everywhere to an action of $SL(\mathcal{H})$ because, essentially, gradient-like vector fields are not complete.
What is interesting though, is that if we consider only points in the space of quantum states $\stsp\subset\mathfrak{T}_{1}\subset\obsp^{*}$, then both Hamiltonian and gradient-like vector fields are complete, and, denoting with $\rho$ the density operator uniquely  associated with the quantum state $\omega\in\stsp$, the integral curve $\gamma_{t}$ of $\mathbb{X}_{\mathbf{a}} + \mathbb{Y}_{\mathbf{b}}$ starting at $\rho$ may be written as:

\be
\gamma_{t}(\rho)=\frac{\gr_{t}\,\rho\,\gr^{\dagger}_{t}}{\Tr(\gr_{t}\,\rho\,\gr_{t}^{\dagger})}=\frac{\mathrm{e}^{(\imath\mathbf{a} + \mathbf{b})t}\,\rho\,\mathrm{e}^{(-\imath\mathbf{a} + \mathbf{b})t}}{\Tr(\mathrm{e}^{(\imath\mathbf{a} + \mathbf{b})t}\,\rho\,\mathrm{e}^{(-\imath\mathbf{a} + \mathbf{b})t})}\,.
\ee
These vector fields provide us with an action of $SL(\mathcal{H})$ on the space of quantum states $\stsp$, and this action is such that $\stsp$ is partitioned in the disjoint union of orbits of $SL(\mathcal{H})$ made up by density operators with fixed rank.\cite{grabowski_kus_marmo-geometry_of_quantum_systems_density_states_and_entanglement, grabowski_kus_marmo-symmetries_group_actions_and_entanglement}

\section{Probability distributions}\label{sec: 3-dim classical case}

Probability distributions on a finite sample $\chi$ space constitute a simplex.
This is a convex body generated by means of convex combinations of a finite numbers of extremal points.
As a matter of fact, it can be shown that the space of states of a $C^{*}$-algebra $\appa_{c}$ is a simplex if and only if the algebra is commutative.
In this setting, the sample space $\chi$ is the so-called Gelfand spectrum of $\appa_{c}$.
We shall consider, as a running example, a sample space of three elements.
Here, a probability distribution is a function:

\be
p\colon \chi=\{1,\,2,\,3\}\rightarrow [0,\,1]\subset\mathbb{R}
\ee
such that $p(1)=p_{1}\geq0$, $p(2)=p_{2}\geq0$, $p(3)=p_{3}\geq0$ and:

\be
p_{1} + p_{2} + p_{3} = 1\,.
\ee
The family of probability distributions, say $\mathcal{P}(\chi)$, may be represented as a convex subset $\Sigma$ of a topological vector space $E$.
In the present case it would be the real vector space $\mathbb{R}^{3}$, subspace of the complex vector space $\mathbb{C}^{3}$.
It is clear that $\{p_{j}\}_{j=1,2,3}$ is a Cartesian coordinate system on $E$, so that the condition $p_{1} + p_{2} + p_{3} = 1$ defines an affine hyperplane in the vector space $E$.

\vsp

A linear transformation on $E$ mapping $\Sigma\rightarrow\Sigma$ may be characterized by means of dual points in $E^{*}$.
Specifically, we consider the element $(1,1,1)\in E^{*}$, and require that the linear map $\mathbf{A}$ satisfies $(1,1,1)\,\mathbf{A}=(1,1,1)$.
Writing $\mathbf{A}=\left(\begin{matrix} a_{1} & b_{1} & c_{1} \\ a_{2} & b_{2} & c_{2} \\ a_{3} & b_{3} & c_{3}\end{matrix} \right)$, we obtain the conditions:

\be
\begin{split}
a_{1} + a_{2} + a_{3} &=1 \\
b_{1} + b_{2} + b_{3} &=1 \\
c_{1} + c_{2} + c_{3} &=1\,.
\end{split}
\ee
from which it follows that $\mathbf{A}$ is a pseudo-stochastic matrix.\cite{chruscinski_manko_marmo_ventriglia-on_pseudo-stochastic_matrices_and_pseudo-positive_maps, manko_marmo_simoni_ventriglia-semigroup_of_positive_maps_for_qudit_states_and_entanglement_in_tomographic_probability_representation}
If we require $\mathbf{A}$ to be invertible, that is, $\mathrm{det}(\mathbf{A})\neq0$, we obtain a Lie group which maps the affine subspace containing $\Sigma$ into itself.
This group will be a six dimensional Lie group, a subgroup of  $GL(3\,,\mathbb{R})$.
It is isomorphic with $IGL(2\,,\mathbb{R})$.
The algebra of pseudo-stochastic  matrices  may also be obtained as the group-algebra of the subgroup $IGL(2,\mathbb{R})$.
On the other hand, if we require the elements of $\mathbf{A}$ to be non-negative, we obtain a stochastic matrix.
Stochastic matrices themselves will be elements of a convex space, indeed it is easy to see that  that the convex sum of two stochastic matrices is still stochastic.
Invertible stochastic matrices form only a semigroup, because the inverse of a stochastic matrix need not preserve positivity.
If we impose that $\mathbf{A}$ fixes $\left(\begin{matrix} 1 \\ 1 \\ 1\end{matrix}\right)\in E$ together with $(1,1,1)\in E^{*}$, we obtain a bistochastic matrix, that is, a matrix such that its columns and rows represent probability vectors.
It is not difficult to show that the maximal subgroup of bistochastic matrices is the permutation group.
It is now possible to consider the group-algebra of the group of bistochastic matrices and we obtain again the algebra of pseudo-stochastic matrices if we use real coefficients.
This will be the most economical way to obtain the algebra of pseudostochastic matrices.

\vsp

In analogy with the bivector fields of the quantum case, we may wonder about the existence of bivector fields which generate Hamiltonian and gradient vector fields on the simplex.
This is indeed the case as we now show.
Consider the following bivector field on $E$:

\be\label{eqn: antisymmetric bivector field on 3-simplex}
\Lambda=\left(p_{1} + p_{2} + p_{3}\right)\left(\frac{\partial}{\partial p_{1}}\wedge\frac{\partial}{\partial p_{2}} + \frac{\partial}{\partial p_{2}}\wedge\frac{\partial}{\partial p_{3}} + \frac{\partial}{\partial p_{3}}\wedge\frac{\partial}{\partial p_{1}}\right)\,.
\ee
It is easy to check that $\mathcal{C}:=p_{1} + p_{2} + p_{3}$ is a Casimir function in the sense that:

\be
\Lambda(\mathrm{d}\mathcal{C}\,,\mathrm{d}f)=0\;\;\;\;\forall f\in\mathcal{F}(E)\,.
\ee 
Since $\mathcal{C}=p_{1} + p_{2} + p_{3}$ is a Casimir function, Hamiltonian vector fields:

\be\label{eqn: hamiltonian vector fields 3-simplex}
X_{f}=\Lambda(\mathrm{d}f\,,\cdot)
\ee
are tangent to the affine hyperplanes $p_{1} + p_{2} + p_{3}=c$ with $c\in\mathbb{R}$.
Furthermore, as the overall coefficient is a Casimir, and all vector fields entering the definition Eq.  \eqref{eqn: antisymmetric bivector field on 3-simplex} of the  bivector field  $\Lambda$ pairwise commute, it follows that the associated Lie algebra bracket satisfies the Jacobi identity.
Thus, $\Lambda$ gives rise to a Poisson bracket among functions given in the linear coordinates $p_i$ by:

\be
\{p_{1}\,,p_{2}\}=\{p_{2}\,,p_{3}\}=\{p_{3}\,,p_{1}\}=p_{1} + p_{2} + p_{3}\,.
\ee
The symplectic leaves of the Poisson bivector $\Lambda$ are the two-dimensionl affine hyperplanes parallel to $p_{1} + p_{2} + p_{3}=1$  (except the degenerate set $p_1 + p_2 + p_3 = 0$).

\begin{remark}
The bivector field appears to be the sum of bivector fields generating the Heisenberg-Weyl algebra on each two-dimensional subspace $(p_{1}\,,p_{2})$, $(p_{2}\,,p_{3})$ and $(p_{3}\,,p_{1})$ multiplied by the overall Casimir function $\mathcal{C}=p_{1} + p_{2} + p_{3}$.
This is a basic observation to show that a similar construction will provide a Poisson bivector  on any simplex of any finite dimension.
\end{remark}

If $f_{\mathbf{a}}=a^{j}p_{j}$ is a linear function on $E$, the Hamiltonian vector field $X_{\mathbf{a}}$ associated with $f_{\mathbf{a}}$ by means of $\Lambda$ reads:

\be
X_{\mathbf{a}}=\mathcal{C}\,\left(a^{1}\,\left(\frac{\partial}{\partial p_{2}} - \frac{\partial}{\partial p_{3}}\right) + a^{2}\,\left(\frac{\partial}{\partial p_{3}} - \frac{\partial}{\partial p_{1}}\right) + a^{3}\,\left(\frac{\partial}{\partial p_{1}}- \frac{\partial}{\partial p_{2}}\right)\right)\,.
\ee
Going back to the bivector field $\Lambda$ given by Eq. \eqref{eqn: antisymmetric bivector field on 3-simplex}, by using quadratic functions  on the simplex, we obtain Hamiltonian vector fields which are infinitesimal generators of  the special linear group $SL(2\,,\mathbb{R})$.
This follows because the simplex is two-dimensional and the symplectic form associated to the bivector field on each  symplectic leaf will be a volume form so that symplectic transformations coincide with volume-preserving transformations.

%

Let us introduce the following Cartesian coordinates on $E$:

\be
x_{1}=\frac{1}{2}\left(p_{1} - p_{3}\right)\,,\;\;\; x_{2}=\frac{1}{2}\left(p_{2} - p_{1}\right)\,,\;\;\; x_{3}=p_{1} + p_{2} + p_{3} -1\,.
\ee
The affine subspace containing the simplex $\Sigma$ is then defined by the condition $x_{3}=0$, and thus this coordinate system is analogous to the Cartesian coordinate system introduced in the quantum setting in the sense that in both cases the affine hyperplane containing the convex set we are interested in (space of relevant states) is a coordinate plane\footnote{In the quantum case it is the affine hyperplane defined by $x_{0}=\frac{1}{\sqrt{n}}$, where $n$ is the dimension of the quantum system.}
An explicit calculation shows that the coordinate expression of $\Lambda$ in this new coordinate system is:

\be
\Lambda=\mathcal{C}'\,\left(\frac{\partial}{\partial x_{1}} \wedge\frac{\partial }{\partial x_{2}}\right)\,,
\ee
with $\mathcal{C}'=\frac{3}{4}\left(x_{3} + 1\right)$ a Casimir function.
It is clear that on every two-dimensional affine hyperplane $x_{3}=c$, with $1\neq c\in\mathbb{R}$, $\Lambda$ becomes an invertible Poisson tensor, and its associated symplectic form is precisely the ``canonical'' symplectic form of the plane.
The Hamiltonian vector field $X_{\mathbf{a}}$ associated with the linear function $f_{\mathbf{a}}=a^{j}x_{j}$ is:

\be
X_{\mathbf{a}}=\mathcal{C}'\,\,\left(a^{1}\,\frac{\partial}{\partial x_{2}} - a^{2}\,\frac{\partial}{\partial x_{1}}\right)\,.
\ee
Now, we consider the functions:

\be\label{eqn: quadratic functions for the 3-simplex}
f_{1}=\frac{(x_{1})^{2} + (x_{2})^{2}}{2}\,,\;\;\;f_{2}=\frac{(x_{1})^{2} - (x_{2})^{2}}{2}\,,\;\;\;f_{3}= -x_{1}\,x_{2}\,,
\ee
and we compute their associated Hamiltonian vector fields:

\be
\begin{split}
X_{f_{1}}&=\mathcal{C}'\,\left(x_{1}\,\frac{\partial}{\partial x_{2}} - x_{2}\,\frac{\partial}{\partial x_{1}}\right)\\
&\\
X_{f_{2}}&=\mathcal{C}'\,\left(x_{1}\,\frac{\partial}{\partial x_{2}} + x_{2}\,\frac{\partial}{\partial x_{1}}\right)\\
& \\
X_{f_{3}}&=\mathcal{C}'\,\left(x_{1}\,\frac{\partial}{\partial x_{1}} - x_{2}\,\frac{\partial}{\partial x_{2}}\right)\,.
\end{split}
\ee
A direct computation shows that $X_{f_{1}},\,X_{f_{2}}$ and $X_{f_{3}}$ close on a realization of the Lie algebra of $SL(2\,,\mathbb{R})$ as claimed.
Furthermore, if, in addition to $X_{f_{1}},\,X_{f_{2}}$ and $X_{f_{3}}$, we consider the Hamiltonian vector fields $X_{1}$ and $X_{2}$ associated, respectively, with the coordinate functions $x_{1}$ and $x_{2}$, we obtain a realization of $\mathfrak{isl}(2\,,\mathbb{R})$, that is, the Lie algebra of the inhomogeneous special linear group $ISL(2\,,\mathbb{R})$ which is isomorphic to the Lie algebra of the Lie group  of invertible pseudo-stochastic matrices with unit determinant.\cite{chruscinski_manko_marmo_ventriglia-on_pseudo-stochastic_matrices_and_pseudo-positive_maps, manko_marmo_simoni_ventriglia-semigroup_of_positive_maps_for_qudit_states_and_entanglement_in_tomographic_probability_representation}

These vector fields are complete, and thus the realization of the Lie algebra integrates to an action of the Lie  group $ISL(2\,,R)$.
This action preserves the affine hyperplane in which the simplex $\Sigma$ lives, but it does not preserve the simplex itself, indeed it does not preserve positivity.
Moreover, every affine hyperplane which is parallel to the affine hyperplane $\mathcal{C}=p_{1}+p_{2}+p_{3}=1$, except the one with $\mathcal{C}=0$, turns out to be an orbit of this group action.

\begin{remark}
It would be possible to define a quadratic Poisson Bracket on the simplex by means of the following bivector field:

\be
\Pi = p_{1}\,p_{2}\,\frac{\partial}{\partial p_{1}}\wedge\frac{\partial}{\partial p_{2}} + p_{2}\,p_{3}\,\frac{\partial}{\partial p_{2}}\wedge\frac{\partial}{\partial p_{3}} + p_{3}\,p_{1}\,\frac{\partial}{\partial p_{3}}\wedge\frac{\partial}{\partial p_{1}}\,,
\ee
from which it follows that:

\be
\{p_{1}\,,p_{2}\}=p_{1}\,p_{2}\,,\;\;\;\{p_{2}\,,p_{3}\}=p_{2}\,p_{3}\,,\;\;\;\{p_{3}\,,p_{1}\}=p_{3}\,p_{1}\,.
\ee
Notice that $p_{1}p_{2}p_{3}$ is a Casimir function for $\Pi$.
It would correspond to the determinant of a diagonal matrix associated with the probability vector instead of the trace, as the bracket we are considering.
Hamiltonian vector fields $X_{\mathbf{a}}$ associated with linear functions $f_{\mathbf{a}}=a^{j}p_{j}$ would be nonlinear and generators of a nonlinear action:

\be
X_{\mathbf{a}}=\left(a^{3}p_{3} - a^{2}p_{2}\right)\,p_{1}\,\frac{\partial}{\partial p_{1}} + \left(a^{1}p_{1} - a^{3}p_{3}\right)\,p_{2}\,\frac{\partial}{\partial p_{2}} + \left(a^{2}p_{2} - a^{1}p_{1}\right)\,p_{3}\,\frac{\partial}{\partial p_{3}}\,.
\ee
As they preserve the determinant, if we start with a probability vector, the associated one-parameter group, by continuity, would not take the vector out of the simplex.
However, their action would not respect convex combinations.

Note that, when $p_{j}>0$, we can take the coordinate chart $x_{j}=\ln(p_{j})$ so that we can write $\Pi$ as:

\be
\begin{split}
\Pi &= \frac{\partial}{\partial \ln(p_{1})}\wedge\frac{\partial}{\partial \ln(p_{2})} + \frac{\partial}{\partial \ln(p_{2})}\wedge\frac{\partial}{\partial \ln(p_{3})} + \frac{\partial}{\partial \ln(p_{3})}\wedge\frac{\partial}{\partial \ln(p_{1})}=\\
&= \frac{\partial}{\partial x_{1}}\wedge\frac{\partial}{\partial x_{2}} + \frac{\partial}{\partial x_{2}}\wedge\frac{\partial}{\partial x_{3}} + \frac{\partial}{\partial x_{3}}\wedge\frac{\partial}{\partial x_{1}}\,,
\end{split}
\ee
and:

\be
\{x_{1}\,,x_{2}\}=\{x_{2}\,,x_{3}\}=\{x_{3}\,,x_{1}\}=1\,.
\ee
In this coordinate chart the Fisher-Rao metric tensor reads:

\be
\begin{split}
g&=p_{1}\,\mathrm{d}\ln(p_{1})\otimes\mathrm{d}\ln(p_{1}) + p_{2}\,\mathrm{d}\ln(p_{2})\otimes\mathrm{d}\ln(p_{2}) + p_{3}\,\mathrm{d}\ln(p_{3})\otimes\mathrm{d}\ln(p_{3})=\\
&=\mathrm{e}^{x_{1}}\,\mathrm{d}x_{1}\otimes\mathrm{d}x_{1} + \mathrm{e}^{x_{2}}\,\mathrm{d}x_{2}\otimes\mathrm{d}x_{2} + \mathrm{e}^{x_{3}}\,\mathrm{d}x_{3}\otimes\mathrm{d}x_{3}\,.
\end{split}
\ee
\end{remark}

\vsp

A symmetric tensor $G$ on $E$  may be constructed following the same analogy:

\be\label{eqn: symmetric bivector field on 3-simplex}
\begin{split}
G&=2\left(p_{1} + p_{2} + p_{3}\right)\left(\frac{\partial}{\partial p_{1}}\otimes\frac{\partial}{\partial p_{1}} + \frac{\partial}{\partial p_{2}}\otimes\frac{\partial}{\partial p_{2}} + \frac{\partial}{\partial p_{3}}\otimes\frac{\partial}{\partial p_{3}}\right) - \\
& \\
&- \left(p_{1} + p_{2} + p_{3}\right)\left(\frac{\partial}{\partial p_{1}}\otimes_{S}\frac{\partial}{\partial p_{2}} + \frac{\partial}{\partial p_{2}}\otimes_{S}\frac{\partial}{\partial p_{3}} + \frac{\partial}{\partial p_{3}}\otimes_{S}\frac{\partial}{\partial p_{1}}\right) \,.
\end{split}
\ee
A direct computation shows that $\mathcal{C}=p_{1} + p_{2} + p_{3}$ is again a ``Casimir'' in the sense that:

\be
G\left(\mathrm{d}\mathcal{C}\,,\mathrm{d}f\right)=0\;\;\;\forall f\in\mathcal{F}(E)\,.
\ee
Consequently, gradient-like vector fields, associated with every smooth function $f\in\mathcal{F}(E)$ by setting:

\be\label{eqn: gradient-like vector fields 3-simplex}
Y_{f}:=G\left(\mathrm{d}f\,,\cdot\right)\,,
\ee
turn out to be tangent to the family of affine hyperplanes defined by the condition $\mathcal{C}=c$ with $c\in\mathbb{R}$.

It is not difficult to show that all Hamiltonian vector fields associated with linear functions turn out to be ``Killing'' vector fields for the symmetric tensor $G$, that is:

\be
L_{X_{\mathbf{a}}}\,G=0\,,
\ee
where $L$ denotes the Lie derivative.

If we consider a linear function $f_{\mathbf{a}}=a^{j}p_{j}$, the associated gradient-like vector field is:

{\footnotesize \be
Y_{\mathbf{a}}=\mathcal{C}\left(a^{1}\left(2\frac{\partial}{\partial p_{1}} - \frac{\partial}{\partial p_{2}} - \frac{\partial}{\partial p_{3}}\right) + a^{2}\left(2\frac{\partial}{\partial p_{2}} - \frac{\partial}{\partial p_{3}} - \frac{\partial}{\partial p_{1}}\right) + a^{3}\left(2\frac{\partial}{\partial p_{3}} - \frac{\partial}{\partial p_{1}} - \frac{\partial}{\partial p_{2}}\right)\right)\,.
\ee}
Now, since $\mathcal{C}$ is a Casimir for both $\Lambda$ and $G$, it is then easy to prove that:

\be
\left[X_{\mathbf{a}}\,,X_{\mathbf{b}}\right]=\left[X_{\mathbf{a}}\,,Y_{\mathbf{b}}\right]=\left[Y_{\mathbf{a}}\,,Y_{\mathbf{b}}\right]=0
\ee
for every Hamiltonian vector field $X_{\mathbf{a}},X_{\mathbf{b}}$ and every gradient-like vector field $Y_{\mathbf{a}},Y_{\mathbf{b}}$.

\vsp

We can use $\Lambda$ and $G$ to define, respectively, a Lie product $[[\cdot\,,\cdot]]$ and a symmetric product $\odot$ on linear functions:

\be
[[f_{\mathbf{a}}\,,f_{\mathbf{b}}]]:=\Lambda\left(\mathrm{d}f_{\mathbf{a}}\,,\mathrm{d}f_{\mathbf{b}}\right)\,,
\ee
\be
f_{\mathbf{a}}\odot f_{\mathbf{b}}:=G\left(\mathrm{d}f_{\mathbf{a}}\,,\mathrm{d}f_{\mathbf{b}}\right)\,.
\ee
The structure constants of these products are easily computed on the basis given by $e_{1}:=\frac{p_{1} - p_{2}}{\sqrt{3}},\,e_{2}:=\frac{p_{2}-p_{3}}{\sqrt{3}},\,e_{3}:=\mathcal{C}=p_{1} + p_{2} + p_{3}$:

\be
[[e_{1}\,,e_{2}]]=e_{3}\,,\;\;\;\;[[e_{1}\,,e_{3}]]=[[e_{2}\,,e_{3}]]=0\,,
\ee
from which it follows that the Lie product $[[\,,]]$ gives a realization of the Heisenberg algebra, and

\be
e_{1}\odot e_{1}=e_{2}\odot e_{2}=2e_{3}\,,\;\;\;e_{1}\odot e_{2}=-e_{3}\,,\;\;\;e_{3}\odot e_{j}=0\;\;\;\forall j\,.
\ee

It is easy to see that the symmetric product $\odot$ is a Jordan product, that is, it is commutative and: \cite{chu-jordan_structures_in_geometry_and_analysis, jordan_vonneumann_wigner-on_an_algebraic_generalization_of_the_quantum_mechanical_formalism}

\be
\left(f_{\mathbf{a}}\odot f_{\mathbf{b}}\right)\odot\left(f_{\mathbf{a}}\odot f_{\mathbf{a}}\right)=f_{\mathbf{a}}\odot \left(f_{\mathbf{b}}\odot\left(f_{\mathbf{a}}\odot f_{\mathbf{a}}\right)\right)\,.
\ee
Next, recalling that $[[e_{3}\,,f_{\mathbf{a}}]]=e_{3}\odot f_{\mathbf{a}}=0$ for all $f_{\mathbf{a}}$, a direct calculation shows that:

\be
0=[[f_{\mathbf{a}}\,,f_{\mathbf{b}}\odot f_{\mathbf{c}}]]= [[f_{\mathbf{a}}\,,f_{\mathbf{b}}]]\odot f_{\mathbf{c}} + f_{\mathbf{b}}\odot[[f_{\mathbf{a}}\,,f_{\mathbf{c}}]]=0\,,
\ee
which means that $[[f_{\mathbf{a}}\,,\cdot]]$ is (trivially) a derivation of $\odot$ for all $f_{\mathbf{a}}$, and we have:

\be
0=(f_{\mathbf{a}}\odot f_{\mathbf{b}})\odot f_{\mathbf{c}} - f_{\mathbf{a}}\odot (f_{\mathbf{b}}\odot f_{\mathbf{c}}) = [[f_{\mathbf{b}}\,,[[f_{\mathbf{c}}\,,f_{\mathbf{a}}]]\,]]=0\,,
\ee
which means that the ``associator'' of the Jordan product is (trivially) related to the Lie product. 
Consequently, the vector space of linear functions endowed with the Lie product $[[\cdot\,,\cdot]]$ and the symmetric product $\odot$ is a Lie-Jordan algebra.

Finally, we may introduce an associative product on the complexification of the space of linear functions, which turns out to define a $C^*$-algebra structure, setting:\cite{falceto_ferro_ibort_marmo-reduction_of_lie-jordan_algebras_and_quantum_states}

\be
f_{\mathbf{a}}\star f_{\mathbf{b}}:=\frac{1}{2}\left(f_{\mathbf{a}}\odot f_{\mathbf{b}} + \imath [[f_{\mathbf{a}}\,,f_{\mathbf{b}}]]\right)\,.
\ee
The associativity of the product follows from the fact that $[[f_{\mathbf{a}}\,,\cdot]]$ is (trivially) a derivation of the Jordan product $\odot$.
The structure constants of this product on the basis $\{e_{j}\}_{j=1,2,3}$ introduced before are:

\be
\begin{split}
e_{1}\star e_{1}&=e_{2}\star e_{2} = e_{3}\,,\;\;\;e_{1}\star e_{2}=\frac{-e_{3} + \imath e_{3}}{2}\,,\\
&\\
e_{2}\star e_{1}&=-\frac{e_{3} + \imath e_{3}}{2}\,,\;\;\;e_{3}\star e_{j}=e_{j}\star e_{3}=0\;\;\;\forall j=1,2,3\,.
\end{split}
\ee
It is clear that there is no unity element in this algebra.

\section{Discussion and conclusions}

In the quantum case briefly resumed in section \ref{sec: quantum}, we started with the Lie-Jordan algebra structure on the space $\obsp$ of observables of the theory.
This allows us to define two bivector fields on the dual space $\obsp^{*}$ in which the space $\stsp$ of quantum states lives by means of equations \eqref{eqn: tildelambda 1} and \eqref{eqn: G 1}.
The bivector field $\Lambda$ associated with the Lie product is skewsymmetric, while the bivector field $G$ associated with the Jordan product is symmetric.
By means of these tensor fields we are able to realize the Lie-Jordan structure of $\obsp$ on the space of linear function on $\obsp^{*}$.
Furthermore, we can associate to every $\mathbf{a}\in\obsp$ a Hamiltonian and a gradient-like vector field setting using, respectively, equations \eqref{eqn: hamiltonian vector fields tildelambda} and \eqref{enq: gradient-like vector fields G}.
These vector fields close a realization of the Lie algebra of the complex general linear group $GL(\mathcal{H})$ on $\obsp^{*}$.
This realization can be integrated to an action of the Lie group $GL(\mathcal{H})$.
Hamiltonian vector fields are tangent to the affine hyperplane $\mathfrak{T}_{1}$ of which $\stsp$ is a convex subset, while gradient-like vector fields are not.
Consequently, we decided to deform the bivector field $G$ into the bivector field $\mathcal{R}$ given by equation \eqref{eqn: R} so that the new gradient-like vector fields  are tangent to the affine hyperplane $\mathfrak{T}_{1}$.
It turns out that these new gradient-like vector fields together with the Hamiltonian vector fields close on a realization of the Lie algebra of the complex special linear group $SL(\mathcal{H})$.
This action does not integrate everywhere to an action of $SL(\mathcal{H})$ (the gradient-like vector fields are not complete), however, it integrates on the subset $\stsp$ of the space of quantum states, and we obtain a partition of $\stsp$ into orbits of $SL(\mathcal{H})$ consisting of quantum states with a given fixed rank.\cite{ciaglia_dicosmo_ibort_laudato_marmo-dynamical_vector_fields_on_the_manifold_of_quantum_states, grabowski_kus_marmo-geometry_of_quantum_systems_density_states_and_entanglement, grabowski_kus_marmo-symmetries_group_actions_and_entanglement}
By using only Hamiltonian and gradient vector fields it is not possible to write Markovian dynamics, and one has to introduce Kraus vector fields. \cite{ciaglia_dicosmo_laudato_marmo-differential_calculus_on_manifolds_with_boundary.applications, ciaglia_dicosmo_ibort_laudato_marmo-dynamical_vector_fields_on_the_manifold_of_quantum_states, carinena_clemente-gallardo_jover-galtier_marmo-tensorial_dynamics_on_the_space_of_quantum_states}
This will not be the case for dynamics on the simplex, marking an important difference between ``classical'' and ``quantum'' evolutions.

\vsp

In classicl information geometry it is customary to look at the space of classical probability distributions as a Riemannian manifold endowed with the Fisher-Rao metric tensor.
Here we wanted to show that this is not the only natural structure available on the simplex.
Guided by the geometry of the quantum case, in section \ref{sec: 3-dim classical case} we introduced a couple of bivector fields on the  classical case of the $3$-dimensional simplex.
We decided not to use the Lie-Jordan algebra structure of the space of classical observables.
Indeed, since the $C^{*}$-algebra of a classical system is commutative, the antisymmetric bivector field associated with the Lie structure would have been identically zero.
We thus proposed the skewsymmetric bivector field $\Lambda$ of equation \eqref{eqn: antisymmetric bivector field on 3-simplex}, and the symmetric bivector field $G$ of equation \eqref{eqn: symmetric bivector field on 3-simplex}.
These bivector fields allow us to define, respectively, Hamiltonian and gradient-like vector fields (see equations \eqref{eqn: hamiltonian vector fields 3-simplex} and \eqref{eqn: gradient-like vector fields 3-simplex}) on the vector space $E$ in which the $3$-simplex lives.
It turns out that Hamiltonian and gradient-like vector fields are tangent to the family of affine hyperplanes in $E$ which are parallel to the affine hyperplane $p_{1} + p_{2} + p_{3}=1$ of which  the $3$-simplex is a convex body.
This is in contrast with the quantum case, where Hamiltonian and gradient-like vector fields are tangent only to the affine hyperplane $\mathfrak{T}_{1}$ containing the space $\stsp$ of quantum states.
On the other hand, the Hamiltonian and gradient-like vector fields associated with linear functions close on a realization of a commutative Lie algebra.
In contrast with the quantum case, the flow of these vector fields, even if is tangent to the affine hyperplane $p_{1} + p_{2} + p_{3}=1$, does not preserve the $3$-simplex.

Interestingly, the symplectic foliation of  $\Lambda$ is such that the leaves are the $2$-dimensional affine hyperplanes parallel to $p_{1} + p_{2} + p_{3}=1$.
On these leaves, except that defined by $p_{1} + p_{2} + p_{3}=0$, the Poisson tensor $\Lambda$ becomes an invertible tensor, and the associated symplectic form is a multiple of the canonical symplectic form associated with the Heisenberg algebra $\mathbb{H}_{3}$.
Then, considering quadratic functions on $E$, we are able to build a realization, in terms of Hamiltonian vector fields, of the Lie algebra $\mathfrak{isl}(2,\mathbb{R})$ of the inhomogeneous real special linear group $ISL(2,\mathbb{R})$ (see paragraph below equation \eqref{eqn: quadratic functions for the 3-simplex}).
According to Refs.\cite{chruscinski_manko_marmo_ventriglia-on_pseudo-stochastic_matrices_and_pseudo-positive_maps, manko_marmo_simoni_ventriglia-semigroup_of_positive_maps_for_qudit_states_and_entanglement_in_tomographic_probability_representation}, this Lie group is the Lie group of pseudo-stochastic matrices.
This is a Lie group acting on $E$ and preserving the affine hyperplane $p_{1} + p_{2} + p_{3}=1$ of which the $3$-simplex $\Sigma$ is a convex body.
Clearly, since the Hamiltonian vector fields are tangent to the hyperplane $p_{1} + p_{2} + p_{3}=1$, the Hamiltonian vector fields of the realization of $\mathfrak{isl}(2,\mathbb{R})$  act transitively on this hyperplane.
This is quite different from the quantum case where Hamiltonian vector fields generate the orbits of the unitary group and therefore are not transitive.

\vsp

We can generalize the construction of the symmetric and antisymmetric tensors on the $3$-simplex to an arbitrary number $n\geq 3$ of dimensions as follows.
We consider the $n$-dimensional vector space $E_{n}$ in which the $n$-dimensional simplex $\Sigma_{n}$ is a convex body.
We introduce a Cartesian coordinate system  $\{p_{j}\}_{j=1,...,n}$ such that $\Sigma_{n}$ is defined by the condition:

\be
\mathcal{C}:=\sum_{j=1}^{n}\,p_{j}=1\,.
\ee
Then, we consider the vector fields:

\be
X_{12}=\frac{\partial}{\partial p_{1}} - \frac{\partial}{\partial p_{2}}\,,\;\; X_{23}=\frac{\partial}{\partial p_{2}} - \frac{\partial}{\partial p_{3}}\,,\;\;\cdots\;\;X_{n1}=\frac{\partial}{\partial p_{n}} - \frac{\partial}{\partial p_{1}}\,,
\ee
and construct the tensor fields:

\be
\Lambda_{n}=\mathcal{C}\,\left(\sum_{(jk),(kl)}\,X_{jk}\wedge X_{kl}\right)\,,
\ee

\be
G_{n}=\mathcal{C}\,\left(\sum_{(jk),(kl)}\,X_{jk}\otimes X_{kl}\right)\,.
\ee
It is clear that $\Lambda_{n}$ is antisymmetric, while $G_{n}$ is symmetric.
Furthermore, being $X_{jk}\left(\mathcal{C}\right)=0$, it is immediate to conclude that Hamiltonian and gradient vector fields, defined using, respectively, $\Lambda_{n}$ and $G_{n}$, are tangent to the affine hyperplanes defined by $\mathcal{C}=c$ with $c\in\mathbb{R}$.
It is not hard to see that, again because $\mathcal{C}$ is a Casimir, the Hamiltonian and gradient-like vector fields associated with linear functions close on a realization of a commutative algebra.
It will be matter of future investigations to understand if it is possible to build a realization of the Lie algebra $\mathfrak{isl}(n,\mathbb{R})$ in terms of Hamiltonian vector fields associated with quadratic functions just as in the $3$-dimensional case.

At the level of Lie algebras it is possible to define a ``contraction procedure'' which takes from the algebra of the unitary group to the Heisenberg-Weyl Lie algebra of the classical case.
Consider the function $x_{1}^{2} + x_{2}^{2} + x_{3}^{2} +
2(1-\lambda)\left[x_{1}x_{2}+x_{2}x_{3}+x_{3}x_{1}\right]$.
The Poisson bivector associated with this Casimir function coincides with the one of $su(2)$ for $\lambda=1$ and for $\lambda =0$ with the Heisenberg-Weyl Lie algebra.
The spherical cup passing through the extremal states of the simplex  becomes more and more flat when $\lambda$ goes to zero.
It is conceivable that by using the procedure of algebra contractions for associative algebras along the lines of  Refs.\cite{alipour_chruscinski_facchi_marmo_pascazio_rezakhani-dynamical_algebra_of_observables_in_dissipative_quantum_systems, chruscinski_facchi_marmo_pascazio-the_observables_of_a_dissipative_quantum_system, ibort_manko_marmo_simoni_stornaiolo_ventriglia-the_quantum_to_classical_transition:contraction_of_associative_products}
 we might be able to go from the special linear group of the noncommutative  case to the group of the commutative case, in some sense to obtain pseudo-stochastic maps out of the pseudopositive maps as examined in Ref.\cite{chruscinski_manko_marmo_ventriglia-on_pseudo-stochastic_matrices_and_pseudo-positive_maps}.
We shall deal with these questions elsewhere.

We conclude by saying that at the moment we cannot present a clear interpretation of the new geometrical structures available on the simplex, but, because they are natural, we feel they must have an interpretation in terms of statistical concepts.

\section{Acknowledgements}

A. I. was partially supported by the Community of Madrid project QUITEMAD+, S2013/ICE-2801, and MINECO grant MTM2014-54692-P.\\
G. Marmo would like to acknowledge the support provided by the Banco de Santander-UCIIIM ``Chairs of Excellence'' Programme 2016-2017.

\end{document}